\documentclass[10pt,prb,twocolumn]{revtex4}

\usepackage{amssymb, amsmath, bm}
\usepackage{verbatim} 

\usepackage{color}
\definecolor{LinkColor} {rgb} {0, 0, .8}
\usepackage[colorlinks=true, linkcolor=LinkColor, citecolor=black, urlcolor=blue,
    pdfpagemode=none,
    pdftitle={Synchronization},
    pdfauthor={Alan Macdonald}] {hyperref}

\hypersetup{pdfstartview=XYZ
    \hypercalcbp{\oddsidemargin+1in}
    \hypercalcbp{\paperheight-\topmargin-1in-\headheight-\headsep}
     1.25}

\begin {document}
%

\title{Comment on: \\
``A one-way speed of light experiment'' \\
by  E. D. Greaves, et al.\\
Am. J. Phys. \textbf{77} (10), 894-896 (2009)}

\author{Alan Macdonald}
\homepage{http://faculty.luther.edu/\textasciitilde macdonal}
\affiliation{Department of Mathematics\\ Luther College\\ Decorah,
IA 52101,  U.S.A. \\ macdonal@luther.edu}

\author{Ettore Minguzzi}
\affiliation{Dipartimento di Matematica Applicata \\
Universit\`a degli Studi di Firenze \\  Via S. Marta 3,  I-50139
Firenze, Italy. \\ ettore.minguzzi@unifi.it}



\maketitle


We assume that the reader is familiar with the paper of Greaves et al.
Our purpose here is to argue that the paper gives no reason to believe
that the authors have measured the one-way speed of light.

The universal time of Newton does not exist.
In particular, if a signal leaves place $A$ in the rest frame of their experiment and arrives at different place $B$,
then the concept of a ``one-way time'' for the trip is not given a priori.
``One-way time'' in the frame \emph{has no meaning} until it is defined.

The standard way to do this is to first define ``Einstein synchronized clocks'' in the frame of the experiment.
Then the one-way time from $A$ to $B$ is defined as $t_B - t_A$,
where $t_A$ is the departure time according to the clock at $A$
and $t_B$ is the arrival time according to the clock at $B$.

The relationship between the Einstein's light speed postulate, Einstein synchronization,
and the one-way speed of light is subtle.
We have given careful analyses in \cite{macdonald83} and \cite{minguzzi02d}.
The only point that we insist upon here is that ``one-way time'' must be defined before it is used.

To drive this point home, consider another time not given a priori: the ``time between two events''.
This time cannot be defined.
Only the time between two events in an inertial frame can be defined,
and this time is different in different inertial frames.

Turning now to the paper, Greaves et al. write:
\begin{quote}
The measured phase difference will be due to the time delay (time of flight)
in traversing the light path from the source to the sensor,
plus the fixed delay in the long cable and detection electronics.
\end{quote}
Their ``time ... from the source to the sensor'' is a one-way time.
They do not define it.
Call it $t$.

To assume that there is a one-way time $t$ is to assume that ``one-way time'' is defined.
It can be defined in terms of synchronized clocks, as above.
If they have in mind Einstein synchronized clocks, then they beg the question.
For with Einstein synchronization the one-way speed of light is equal to the two-way speed.\cite{macdonald83}
Thus the one way speed with this synchronization has already  been measured to high accuracy.

Their ``fixed delay in the long cable'' is also a one-way time.
They do not define it.
Call it $t_0$.

It is possible to measure a two-way time in the cable:
send a signal from one-end to the other, reflect it back, and measure the elapsed time.
A two-way time needs neither a definition nor synchronized clocks, as it is measured with a single clock.
We suppose that it can be confirmed experimentally that the two-way time is unchanged
if the sensor is relocated in space.

However, to assume that there is a fixed one-way time $t_0$, for example half of the two-way time,
is to assume that ``one-way time'' is defined.

The experiment measured the relationship between the distance $L$
that the light travels from the source to the sensor \cite{L}, and
the round trip time $T$ for the signal to arrive at the detector
(their Figure 1a). They found that $dL/dT = c$, the accepted value
of the speed of light (their Figure 1b). Their only statement that
this measures the one-way speed of light is this:

\begin{quote}
A graph of the measured times as a function of the total distance traveled by the light beam
should be a straight line with slope equal to the speed of light.
When the sensor reaches the position of the mirror, the light has traveled a one-way distance toward the sensor.
The measurements of these delay times and the distances traveled should give the one-way speed of light.
\end{quote}

Since they do not elaborate,
we can only guess why they believe that ``the measurements ... give the one-way speed of light''.
Perhaps they separate the round trip time into the time for light to reach the sensor
plus the fixed delay, as in the first quoted paragraph: $T = t + t_0$.
Then it is easy to interpret the experiment as measuring the one-way speed $c_1$ of light.
For then $T = L/c_1 + t_0$.
With a slightly different $L$, $T + dT = (L + dL)/c_1 + t_0$.
Subtract and rearrange to obtain $dL/dT = c_1$.

However, the one-way times $t$ and $t_0$ have not been defined.
Moreover, since $c_1$ is defined by $c_1 = L/t$, the quantity that they claim to measure has not even been defined.
It is a logical error to use $t, t_0$, or $c_1$.

We see no way to interpret the experiment as measuring the one-way
speed of light.


\begin{thebibliography}{bb}
\bibitem{L}
The authors take $L$ to be the straight line distance from the sensor to the detector.
Our $L$ is the length of ``the light path from the source to the sensor''
from the quote above.
\end{thebibliography}
\end{document}